\documentclass[10pt,aps,prl,amssymb,amsmath]{revtex4}

\newcommand{\ket}{\rangle}

\begin{document}

%
%

\title{The L\"uders Postulate and the Distinguishability of Observables}

\author{Bernhard K. Meister
}

\address{Physics Department, Renmin University of China, Beijing
100872, China\\
}

\begin{abstract}
The L\"uders postulate is reviewed and implications for the
distinguishability of observables  are discussed. As an example the
distinguishability of two similar observables for spin-$\frac{1}{2}$
particles is described. Implementation issues are briefly analyzed.
\end{abstract}







\maketitle

\section{Introduction}

In this paper the L\"uders postulate\cite{luders} 
is used to distinguish similar observables.
 The L\"uders postulate was introduced as a modification of the original
measurement theory of quantum
mechanics as presented by von Neumann\cite{neumann}. 
   It describes unambiguously 
   the measurement
   of observables
with a degenerate spectrum. This paper emphasizes conceptual issues.
Extensions to quantum algorithm have been discussed
elsewhere\cite{meme}.
 In the next Section
 the L\"uders postulate is reviewed and then in the main Section of
 the paper
 with the
 help of a simple example some implications of the postulate are clarified.
In the remaining two Sections
implementation issues are analyzed and 
some concluding remarks are added.





\section{ The L\"uders Postulate}
\label{sec:2}

The L\"uders postulate~\cite{luders,isham}, part of the standard
canon of quantum mechanics, describes the measurement process of
observables with a degenerate spectrum. In the case of operators
with a degenerate spectrum it postulates that the projection of the
initial wave function is onto exactly one point in each degenerate
subspace. The point chosen is the element of the degenerate subspace
`closest' - in terms of transition probability - to the initial wave
function.
This `refinement' of von Neumann's projection postulate 
seems reasonable, since the L\"uders postulate produces measurements
that disturb the wave function minimally. In von Neumann's original
approach the initial wave function is projected onto a full basis,
where the choice of basis depends for the degenerate subspaces on
the nature of the measurement apparatus.
 The
mathematical formulation of the postulate is given next using
standard Dirac notation. We define the normalized eigenfunctions of
the observable $\hat{O}$ with $K$ different eigenvalues, each having
the degeneracy $d_k$, to be $ |\psi_{k,j}\rangle$ , where
 $k=1,2,...,K$
and $j=1,2,...,d_k$. The eigenfunctions allow the definition of the
following set of $K$ projection operators\footnote{Any choice of
orthogonal basis of the degenerate subspace leads to the same
projection operator.}
\begin{eqnarray}
\hat{P}_k=\sum_{j=1}^{d_k}|\psi_{k,j}\rangle \langle \psi_{k,j}| .
\end{eqnarray}
A measurement of an arbitrary pure state
$|\phi\rangle$
now gives according to the
L\"uders postulate the `reduction' to the following states
\begin{eqnarray}
|\phi\rangle \rightarrow {\rm Prob}[O=\lambda_k]^{-1/2} \hat{P}_k
|\phi\rangle
\end{eqnarray}
with the probabilities for the distinct eigenvalues of
\begin{eqnarray}
{\rm Prob}[O=\lambda_k]=\langle \phi | \hat{P}_k|\phi\rangle .
\end{eqnarray}
\section{Distinugishability of similar Observables for Spin-$\frac{1}{2}$ particles}
\label{sec:3}
 The impact of the L\"uders postulate on the
distinguishability of similar observables is next presented. The aim
is to distinguish two known observables, in this section exactly and
in the next section approximately implemented, with the help of the
measurement of a specially selected input wave function. We consider
two observables that measure individual spin-$\frac{1}{2}$
particles. The observable 
is either chosen to be the
identity operator
\begin{eqnarray}
\hat{I} =
\left(
\begin{array}{lr}
1 &  0  \\
0 & 1
\end{array} \right),
\end{eqnarray}
or the operator
\begin{eqnarray}
\hat{J} = 
\left(
\begin{array}{lc}
1 &  0 \\
0 & 1+\delta
\end{array} \right)
\end{eqnarray}
that associates the eigenvalue $1$ to the eigenstate spin-up and
$1+\delta$ to the eigenstate spin-down. We define the states
$|1\rangle$ and $|0\rangle$ to correspond to the two eigenstates of
the observable $\hat{J}$.

 A measurement of either the observable $\hat{I}$ or $\hat{J}$ for an
 input wave function in equal superposition of
spin-up and spin-down, i.e. $1/\sqrt{2}(|1\rangle+ |0\rangle)$, is
carried out initially.   
It gives
for the first observable $\hat{I}$ a direct projection of the wave
function onto itself. For the second observable $\hat{J}$ the
measurement outcome is a mixed state with equal probability in the
state spin-up and spin-down
as long as $\delta$ is nonzero. The unique outcome
$1/\sqrt{2}(|1\rangle+ |0\rangle)$ for the first observable can be
distinguished simply from the mixed state outcome by standard
interference techniques. One can for example, in an additional
apparatus, measure the probability of the wave function in an
appropriate basis like $1/\sqrt{2}(|1\rangle+ |0\rangle)$ and
$1/\sqrt{2}(|1\rangle - |0\rangle)$. In the first case the outcome
will always be $1/\sqrt{2}(|1\rangle+ |0\rangle)$. For the second
case, the mixed state,
 the
probability for each of the basis states is $1/2$. Therefore, the
ability to distinguish the two observables below any chosen error
threshold $\epsilon$ is possible, if sufficient identical copies of
the initial state are prepared and measured in sequence by the same
observable. Each copy available decreases the probability of an
error by a factor of $1/2$. If $m$ copies of the initial state are
prepared, then the probability of an incorrect choice is $2^{-m-1}$.
Therefore, as in the example above, {\it an infinitesimal
deformation of an observable can change a degenerate into
nondegenerate spectrum and can lead to an observable difference}.
The result above is conditional on being able to implement the
observables accurately and being able to carry out the prescribed
measurements efficiently. Implementation issues are discussed next.
 As an aside, the space of Hermitian operators possesses a natural Finslerian
metric~\cite{finsler} permitting a more comprehensive study of
their properties.


\section{Implementation Issues} \label{sec:4}
In this section implementation issues are briefly analyzed. There
are two main challenges that have to be addressed. First, what
implementation accuracy is needed to be able to distinguish the two
observables. Second, how can one estimate the time needed to carry
out a measurement, if the difference between the eigenvalues of an
observable is small. The answer to the second question is dependent
on the interpretation of quantum mechanics one chooses to follow. In
the standard interpretation of quantum mechanics, called the
Copenhagen interpretation closely associated with Bohr and
Heisenberg, measurements are not explicitly modeled and assumed to
be `instantaneous'. This straightforward and simple prescription is
modified in other versions of quantum mechanics. In stochastic
quantum mechanics, for example, the time needed for an energy
measurement is proportional to $\delta H$, i.e. the energy
difference between the different eigenvalues to be distinguished,
where one has to keep in mind that $\delta H$ encompasses both the
system as well as the measurement apparatus and is often
substantial. The discussion of the measurement time in some of the
major interpretations of quantum mechanics is carried out in a
separate paper.  If one assumes the Copenhagen interpretation and
takes measurements to be outside the standard unitary formulation of
quantum mechanics and effectively instantaneous, then the analysis
is drastically simplified.

The implementation accuracy for Hermitian operators is discussed
next. Instead of describing the fine details of what experimentalist
can and cannot do, we instead analyze the impact a finite
implementation accuracy 
has on the distinguishability of observables. It will be shown that
a finite accuracy will reduce the efficiency of being able to
distinguish similar observables only to a limited extent. In the
case of a degenerate observable, any infinitesimal inaccuracy in the
implementation would destroy the degeneracy. A purely random error
would lead to a pair of randomly distributed mutually orthogonal
eigenstates.
 The eigenvectors can be represented in the following way with $(\cos (\alpha) (|1\ket +|0\ket) +\sin(\alpha)(|1\ket - |0\ket))/\sqrt{2}$
 for
 $\alpha$ equally distributed in the interval $[0,\pi/2]$
 corresponding
to
  the first eigenstate and  $(\sin (\alpha)(|1\ket+|0\ket) -\cos(\alpha)(|1\ket - |0\ket))/\sqrt{2}$ corresponding to the
  second eigenstate.
The transition probability for a fixed $\alpha$ to the first basis
state  is $\cos^2(\alpha)$ and to the second basis state is
$\sin^2(\alpha)$. A second measurement, as described in the section
above, in the basis $1/\sqrt{2}(|1\rangle+ |0\rangle)$ and
$1/\sqrt{2}(|1\rangle- |0\rangle)$ will lead to the final
probability of $3/4$, i.e. $\frac{2}{\pi}\int_{0}^{\pi/2} d\alpha
\cos^4(\alpha)+ \sin^4(\alpha)$,
 for the outcome $1/\sqrt{2}(|1\rangle+ |0\rangle)$.

In the other case we assume that the distribution of the eigenstates
is centered around the original basis. In the simplest case we work
with the equivalent of a Gaussian distribution on a circle, i.e. the
von Mises distribution $P(\alpha)\propto \exp(q^2
\cos(\alpha-\alpha_{mean}))$. If $q$ is large then the probability
for a final outcome of $1/\sqrt{2}(|1\rangle+ |0\rangle)$ of the
second measurement is close to $1/2$. A $q$ close to zero leads to a
final probability close to $3/4$. If $q$ is reasonably large, then
the two observables can be distinguished with
limited error probability. 
Here we provided only a schematic analysis to explain the principle
beyond approximate measurements, the details taking into account a
more realistic description of the noise will be given elsewhere.


\section{Conclusion} \label{sec:6}
Let us begin the conclusion by listing two areas that need to be
further investigated:

$\bullet$  What observables can one construct\footnote{ What other
restrictions exist  on measurements besides the Wigner-Araki-Yanase
theorem~\cite{wig,araki} for additive conserved quantities and its
extension to multiplicative conserved quantities? How does one
realize observables with a degenerate spectrum? Are there
construction mechanisms  that are self-stabilizing?}
? How accurately can one construct them?

 $\bullet$  What is the collapse time
associated with particular measurements?

The answer to these questions,  which is strongly tied with the
version of quantum mechanics one adheres to
, determines the distinguishability of similar observables. 
 Eventually, these questions will be settled
 by further theoretical work and more importantly by experiments.
The purpose of this paper was to show that  the L\"uders postulate
has interesting consequences for distinguishing quantum observables,
 and to point out various implementation issues. 
\section*{Acknowledgments}
 This work was presented at and submitted to the
proceedings of the International Conference on the Frontiers of
Nonlinear and Complex Systems in Hong Kong from 24-26 May, 2006. The
financial support of the conference organizers is gratefully
acknowledged. The author wishes to acknowledge numerous stimulating
discussions with Dorje Brody.

\end{document}